\documentclass{ifacconf}
\usepackage{graphicx}
\usepackage{natbib}
\usepackage{amsmath,amssymb,amsfonts}
\usepackage{booktabs}
\usepackage{multirow}
\usepackage{placeins}
\usepackage{float}
\newtheorem{lemma}{Lemma}
\newtheorem{theorem}{Theorem}
\newtheorem{definition}{Definition}
\newtheorem{corollary}{Corollary}
\newtheorem{assumption}{Assumption}
\usepackage{textcomp}
\usepackage[table]{xcolor} 
\usepackage{url} 

\begin{document}
\begin{frontmatter}

\title{Risk Aware Safe Control with Multi-Modal Sensing for Dynamic Obstacle Avoidance\thanksref{footnoteinfo}}
\thanks[footnoteinfo]{This research was supported by the CARMEN+ University Transportation Center, sponsored by the U.S. Department of Transportation under Grant No. 69A3552348327. The views presented are those of the authors and do not necessarily represent the official views of the U.S. Department of Transportation.}

\author[First]{Pei Yu, Chang} 
\author[First]{Qizhe, Xu} 
\author[Second]{Vishnu, Renganathan}
\author[First]{Qadeer, Ahmed}

\address[First]{Department of Mechanical and Aerospace Engineering, The Ohio
State University, Columbus, OH 43212 USA (e-mail: \{chang.2314, xu.5640, ahmed.358\}@osu.edu).}
\address[Second]{Department of Electrical and Computer Engineering, The Ohio State University, Columbus, OH 43212 USA (e-mail: renganathan.5@osu.edu)}

\begin{abstract}                

Safe control in dynamic traffic environments remains a major challenge for autonomous vehicles (AVs), as ego vehicle and obstacle states are inherently affected by sensing noise and estimation uncertainty. However, existing studies have not sufficiently addressed how uncertain multi-modal sensing information can be systematically incorporated into tail-risk-aware safety-critical control. To address this gap, this paper proposes a risk-aware safe control framework that integrates probabilistic state estimation with a conditional value-at-risk (CVaR) control barrier function (CBF) safety filter. obstacle detections from cameras, LiDAR, and vehicle-to-everything (V2X) communication are combined using a Wasserstein barycenter (WB) to obtain a probabilistic state estimate. A model predictive controller generates the nominal control, which is then filtered through a CVaR-CBF quadratic program to enforce risk-aware safety constraints. The approach is evaluated through numerical studies and further validated on a full-scale AV. Results demonstrate improved safety and robustness over a baseline MPC–CBF design, with an average improvement of 12.7\% in success rate across the evaluated scenarios. 

\end{abstract}


\end{frontmatter}

\section{Introduction}

Safe operation in dynamic and uncertain traffic environments is a central requirement for autonomous vehicles (AVs), particularly when interacting with vulnerable road users (VRUs). obstacle-related scenarios are among the most safety-critical cases in autonomous driving due to the limited available reaction time and rapidly changing traffic situation. Reliable control in such cases requires the vehicle to respond quickly while preserving both safety and driving performance. The challenge is further complicated because the ego vehicle and obstacle states available to the controller are not perfectly known in practice.

As a result, safe vehicle control requires explicitly accounting for uncertain state estimates in obstacle-crossing scenarios. Such noisy estimates propagate to downstream control, thereby affecting both collision-risk assessment and the resulting control actions. As illustrated in Fig.\ref{problem_statement}, ignoring such uncertainty will lead to unsafe decisions due to underestimated risk, whereas overly conservative treatment can cause unnecessary braking or degraded driving performance. 

Existing work on autonomous driving has extensively explored the fusion of cameras, LiDAR, and V2X information, primarily to improve localization and perception accuracy (\cite{zhang2023,wang2020,xiang2023}).  More recently, several studies have begun to explicitly model estimation uncertainty. For example, \cite{lou2023} incorporates imprecision into LiDAR-camera-based perception. Despite this progress in uncertainty-aware fusion, integrating state estimates and their associated error structures for downstream safe control remains limited.

CBF remains a powerful paradigm for safety-critical control \cite{panja2024survey}. \cite{ames2019} established the CBF formalism, and \cite{zeng2021} integrated discrete-time CBF with MPC for obstacle avoidance. Classical CBF formulations, however, usually assume exact state information or explicitly bounded uncertainty, making them less reliable in the presence of localization errors, perception noise, and communication delays. To tackle this limitation, \cite{cosner2021measurement} proposed measurement-robust CBF, and \cite{oruganti2023safe} extended this idea to high-order systems. \cite{li2023moving} introduced chance-constrained MPC-CBF for moving-obstacle avoidance under perception uncertainty, while \cite{long2024} proposed sensor-based distributionally robust CBF for safe robot navigation, \cite{ahmadi2022riskcbf} developed CVaR barrier functions to capture tail risk, and \cite{chriat2023} explored Wasserstein distributionally robust CBF using CVaR under distributional uncertainty. Nevertheless, most of these methods are designed and validated mainly on robots or simplified platforms. In our prior work, \cite{renganathan2025enhanced} developed a risk-aware MPC-CVaR-CBF controller, and \cite{chang2025riskbudgeted} proposed a risk-budgeted monitor that switches to CVaR-CBF when safety margins deteriorate. Taken together, the literature still lacks systematic integration between uncertainty-aware multi-modal sensor fusion and tail-risk-aware vehicle control. In particular, how to fuse noisy state estimates from multiple sensing modalities and propagate the resulting uncertainty into a CVaR-based safe control framework remains in its early stages, with little validation on full-scale AV platforms.

\begin{figure}[htbp]
      \centering
      \includegraphics[scale=0.3]{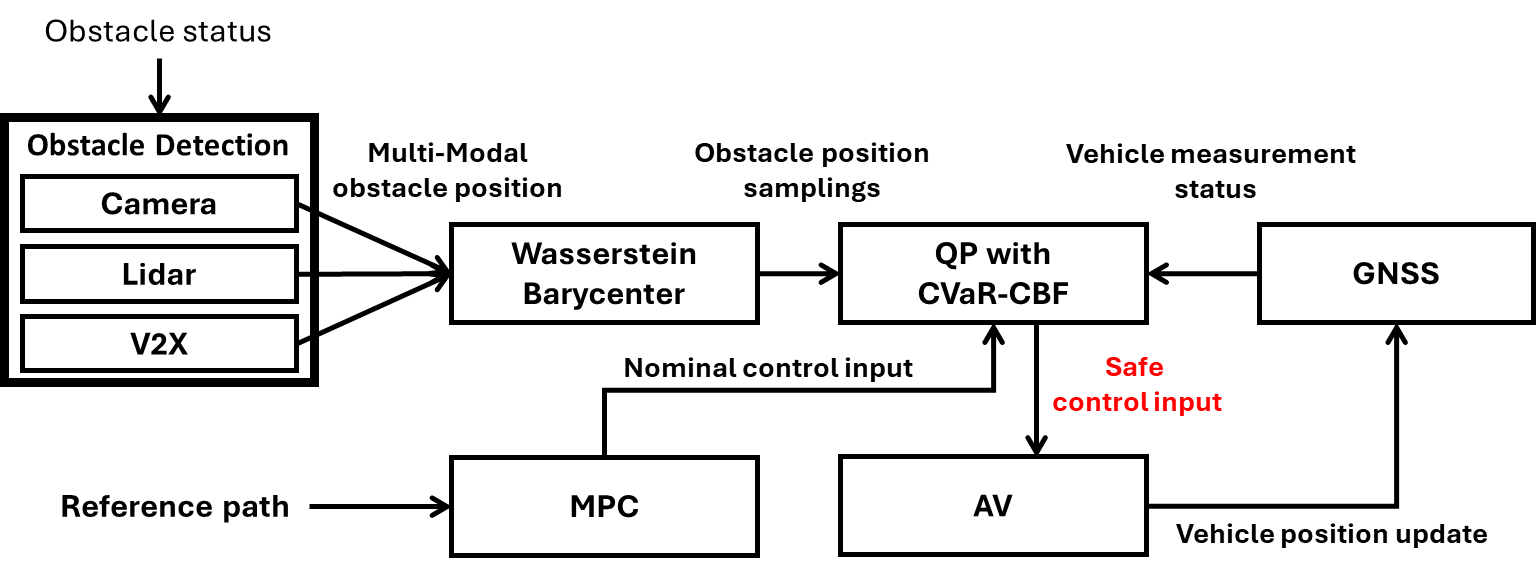}
      \caption{ Multi-source obstacle detections (camera, LiDAR, and V2X) are fused via a Wasserstein barycenter. The nominal MPC control input is filtered by a CVaR-CBF QP to enforce risk-aware safety.}
      \label{proposed_framwork}
\end{figure}

To address this gap, this paper proposes a risk-aware safe control framework that integrates uncertainty-aware multi-modal sensing fusion with CVaR-CBF-based safety filtering (see Fig.\ref{proposed_framwork}). obstacle detections from cameras, LiDAR, and V2X are integrated via a Wasserstein barycenter to obtain a fused state estimate that accounts for sensing errors. A nominal control input is generated by an MPC controller for reference tracking and then filtered through a CVaR-CBF quadratic program to enforce risk-aware safety constraints and obtain the final safe control action. Thus, the proposed framework combines uncertainty-aware fusion, nominal-reference-based control, and tail-risk-sensitive safety assurance within a single control pipeline. 

Specifically, we make the following contributions:

\begin{figure}[t]
      \centering
      \includegraphics[scale=0.3]{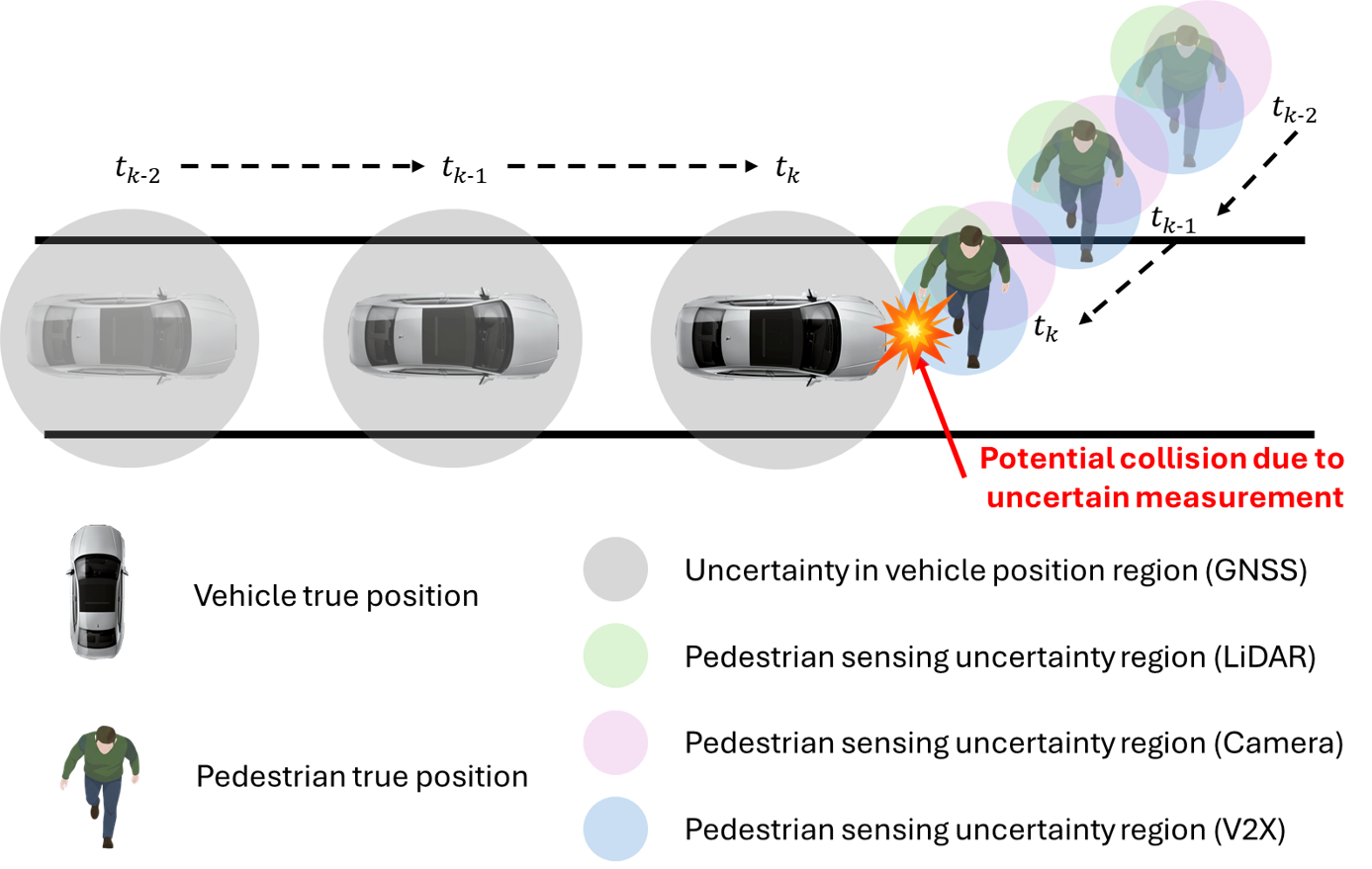}
      \caption{ Illustration of sensing uncertainty in vehicle and obstacle positions. Sensor noise creates a region of possible obstacle locations, some leading to potential collisions, motivating risk-aware safe control.}
      \label{problem_statement}
\end{figure}

\begin{itemize}
    \item We develop a probabilistic multi-modal sensing fusion framework based on the Wasserstein barycenter to obtain a unified obstacle-state estimate under heterogeneous sensor uncertainty.
    
    \item Based on the fused distribution, we formulate a WB-CVaR-CBF safety filter and derive feasibility conditions for the associated optimization problem.
    
    \item We demonstrate real-time deployment on a full-scale autonomous vehicle for sudden obstacle avoidance, and show through simulations and experiments that the proposed method improves robustness and safety over standard CBF baselines under sensing uncertainty.
\end{itemize}

\section{Preliminaries and Problem formulation}



A nonlinear continuous time and control affine system is given by 
\begin{equation} \label{control_affine}
    \dot{\mathbf{x}} = f(\mathbf{x}) + g(\mathbf{x})\mathbf{u}, 
\end{equation}
where $\mathbf{x} \in D \subset \mathbb{R}^n$ and $\mathbf{u} \in \mathcal{U} \subset \mathbb{R}^m$ represent the system state and control input. $\mathcal{U}$ is the set of admissible control inputs. The functions $f : \mathbb{R}^n \rightarrow \mathbb{R}^n, \quad$ and $g : \mathbb{R}^n \rightarrow \mathbb{R}^{n \times m}$ are locally Lipschitz continuous.

\subsection{Safe Sets and Control Barrier Functions} 

For the given control affine system in (\ref{control_affine}), a safe set $\mathcal{C}$ is defined as the superlevel set of a continuously differentiable function $h(\mathbf{x})$, which can be written as:
\begin{equation}
\mathcal{C} = \{ \mathbf{x} \in \mathcal{D} \subset \mathbb{R}^n : h(\mathbf{x}) \geq 0 \}.
\end{equation}
The set $\mathcal{C}$ is rendered forward invariant by a controller $\mathbf{u} \in \mathcal{U}$ if for every $\mathbf{x_0} \in \mathcal{C}$, $\mathbf{x_t} \in \mathcal{C}$ for and all $t > t_0$.

The system is \textit{safe} if we can implement the control input that would render the set $\mathcal{C}$ forward invariant. From  \cite{ames2019, pradeep2024}, for a given set $\mathcal{C}$, $h(\mathbf{x})$ is a \textit{control barrier function (CBF)} with $\frac{\partial h}{\partial (\mathbf{x})}(\mathbf{x}) \neq 0$ if there exists an extended class $\mathcal{K}_\infty$ function $\alpha$ such that for the control system (\ref{control_affine}):
\begin{equation}
\sup_{\mathbf{u} \in \mathcal{U}} \left[ L_f h(\mathbf{x}) + L_g h(\mathbf{x}) \mathbf{u} + \alpha(h(\mathbf{x})) \right] \geq 0, \quad \forall \mathbf{x} \in \mathcal{C}.
\label{eq:cbf}
\end{equation}

$L_f h(\mathbf{x})$ is the is the Lie derivative of $h$ along $f$, and $L_g h(\mathbf{x})$ is the is the Lie derivative of $h$ along $g$.

The formulation in (\ref{cbf_qp}) is referred to as the Control Barrier Function-based Quadratic Program (CBF-QP) that ensures safety.
\begin{equation}
\label{cbf_qp}
\begin{aligned}
\mathbf{u} =\; & \underset{\mathbf{u} \in \mathcal{U} \subset \mathbb{R}^m}{\text{min}} \; \frac{1}{2} \| \mathbf{u} - \mathbf{u}_{\text{nom}} \|^2 \\
\text{s.t. } & L_f h(\mathbf{x}) + L_g h(\mathbf{x}) \mathbf{u} + \alpha\left( h(\mathbf{x}) \right) \geq 0,
\end{aligned}
\end{equation}
where $\mathbf{u}_{\mathrm{nom}}$ denotes the nominal control input generated by the MPC controller. In the proposed architecture, closed-loop stability is primarily attributed to the nominal MPC design, whereas the CBF QP serves as a safety filter that enforces safety constraints through minimal modification of the nominal input.

\subsection{Problem Formulation}

The uncertainties addressed in this paper can be divided into two primary sources: (i) the uncertainty in AV's position. This could be due to input disturbances or the localization measurement uncertainty from the AV's GPS and (ii) the measurement uncertainties from multi-modal sensing through multi-sources. This section explores how uncertainty is modeled. We consider obstacle position measurements obtained from multi-modal sensing through three types of sensors: LiDAR, camera, and V2X. All measurement positions are represented in the East-North-Up (ENU) coordinate system, which serves as the inertial reference frame.

\subsubsection{AV Localization Uncertainty}
We denote the true vehicle's $x, y$ position at time step $k$ as $\mathbf{Z}
_{v,k} = (x_{v,k}, y_{v,k})$, and model the measured position as a Gaussian random variable: 
\begin{equation}
\label{gps_measure}
\hat{\mathbf{Z}}_{v,k} \sim \mathcal{N}(\mathbf{Z}_{v,k}+\mu^{(\text{GPS})}, \sigma^{({\text{GPS}})}),
\end{equation}
where $\hat{\mathbf{Z}}_{v,k} \in \mathbb{R}^2$ is the measured position, $\mu^{(\text{GPS})}$ is the mean deviation and $\sigma^{({\text{GPS}})}$ is a diagonal covariance matrix representing the position uncertainty.

\subsubsection{Multi-Modal State Uncertainty}
The dynamic model of the obstacle is assumed to be as follows:
\begin{equation} \label{obstacle_dynamics}
    \mathbf{o_{s,k+1}} = f_{o}(\mathbf{o_{s,k}}) + \omega_{s,k},
\end{equation}
where \( s \in \{ \text{LiDAR}, \text{Camera}, \text{V2X} \} \), $\mathbf{o_{s,k}} \subset \mathbb{R}^n$ denotes the states of the obstacle at time k, $f_{o}$ is an unknown function, and $\omega_{s,k}$ is disturbance regarding to the sensors. 
The true obstacle position is denoted as \( \mathbf{Z}_{o,k} = (x_{o,k}, y_{o, k}) \in \mathbb{R}^2 \), and the obstacle position measurement from the sensor is denoted as \( \hat{\mathbf{Z}}_{o,k}^{(s)} = (x_{o,k}, y_{o, k}) \in \mathbb{R}^2 \) at time step $k$. 

In this setup, all three sensors perceive the same dynamic obstacle, but each may include different noise characteristics due to various factors. For instance, LiDAR sensors can produce distorted point clouds, which may lead to inaccurate obstacle localization. Cameras are prone to motion blur during fast vehicle movement, degrading image quality and affecting vision-based detection algorithms. V2X communication relies on wireless transmission, which is susceptible to latency and packet loss. All of these reasons can lead to outdated position information, introducing inconsistency in multi-modal sensing.


In order to model the noise distribution, we assume that the measurement uncertainty in each sensor follows a Gaussian distribution on Hilbert spaces, with sensor-specific covariance $\sigma^{(s)}$, and mean bias denoted as $\mu^{(\text{s})}$. Due to V2X latency and network-induced delays, we consider the possibility that the measurements is biased, especially for fast-moving or dynamic obstacles. 
The resulting Gaussian distribution models for the sensors are as follows:
\begin{subequations}
\label{sensor_dis}
\begin{align}
\hat{\mathbf{Z}}_{o,k}^{(\text{LiDAR})} &\sim \mathcal{N}(\mathbf{Z}_{o,k}+\mu^{(\text{LiDAR})}, \sigma^{(\text{LiDAR})}), \\
\hat{\mathbf{Z}}_{o,k}^{(\text{Camera})} &\sim \mathcal{N}(\mathbf{Z}_{o,k}+\mu^{(\text{Camera})}, \sigma^{(\text{Camera})}), \\
\hat{\mathbf{Z}}_{o,k}^{(\text{V2X})} &\sim \mathcal{N}(\mathbf{Z}_{o,k} + \mu^{(\text{V2X})}, \sigma^{(\text{V2X})}).
\end{align}
\end{subequations}

Let $L^2(\mathcal{D})$ denote the space of square-integrable functions mapping from $\mathcal{D}$ to $\mathbb{R}^n$. The covariance function $\sigma^{(s)}$ is associated with an integral operator $\Sigma^{(s)}: L^2(\mathcal{D}) \to L^2(\mathcal{D})$\cite{mallasto2017}. The covariance operators for three sensors can be written as $\Sigma^{(\text{LiDAR})}$, $\Sigma^{(\text{Camera})}$, $\Sigma^{(\text{V2X})}$.

\section{Multi-Modal State Estimation and Wasserstein Barycenter}\label{sec:wb}

This section discusses on how Wasserstein Barycenter is formulated for multi-modal sensing. For defining the Wasserstein metric and Barycenters in the Wasserstein space, consider two vectors $\xi_1$ and $\xi_2$, supported on a set $\Xi \subseteq \mathbb{R}^n$ associated with probability measures $\mu^{(1)}$ and $\mu^{(2)}$, respectively. Let $\mathcal{P}_p(\Xi) \subseteq \mathcal{P}(\Xi)$ be the Borel probability measures over $\Xi$ with finite $p$-order moments. With $p \geq 1$, the $p-$Wasserstein distance between the probability measures $\mu^{(1)}$ and $\mu^{(2)} \in \mathcal{P}_p(\Xi)$ is given by 
{\small
\begin{equation} \label{p-W-dist}
    W_p(\mu^{(1)}, \mu^{(2)}):= \left( \inf_{\pi \in \Pi(\mu^{(1)}, \mu^{(2)})} \int_{\Xi \times \Xi} \| \xi_1-\xi_2 \|^p  \mathrm{d}\pi(\xi_1,\xi_2) \right)^{\frac{1}{p}},
\end{equation}
}

where $\|.\|$ is the Euclidean norm on $\mathbb{R}^n$, $\Pi(\mu^{(1)}, \mu^{(2)})$ is the set of joint distributions on $\Xi \times \Xi$ with marginals $\mu^{(1)}$ and $\mu^{(2)}$. 
In this work, we adopt $p=2$. This choice is motivated by the Gaussian sensor models considered in this work, for which the 2-Wasserstein barycenter admits a unique solution with a tractable mean and covariance characterization.
Intuitively, consider $\mu^{(1)}$ and $\mu^{(2)}$ to measure the same quantity using two different sensors. However, with respect to our problem formulation, we have three sensors (camera, LiDAR, and V2X) measuring the same obstacle, following three distributions - $(\mu^{(\text{LiDAR})}, \mu^{(\text{Camera})}, \mu^{(\text{V2X})})$. Thus, the mean of the three distributions can be determined through the Wasserstein Barycenter \cite{agueh2011barycenters}.

From \cite{agueh2011barycenters}, the $\lambda$-weighted empirical 2-Wasserstein Barycenter for finite set of probability measures $\{\mu^{(1)}, \mu^{(2)}, ..., \mu^{(N)}\}$ with second moments is defined by 
\begin{equation} \label{WB}
    \mu^{(\text{WB})} = \inf_{\mu} \sum_{s=1}^{N} \lambda_s W_2^2(\mu, \mu^{(s)}),
\end{equation}
where $\lambda_s$ are positive weights such that $\sum_{s=1}^{N} \lambda_s =1$. In our current formulation, the 2-Wasserstein Barycenter is considered for its existence and uniqueness properties. 

\subsection{Example on multi-modal sensor distribution}
In this paper, each sensor measurement distribution is modeled as Gaussian. 
With positive barycentric weights $\{\lambda_s\}_{s=1}^{N}$ satisfying $\sum_{s=1}^{N}\lambda_s = 1$, the corresponding 2-Wasserstein barycenter is well defined and unique.

\begin{lemma} 
\label{wb_unique}
Let $\{\hat{\mathbf{Z}}_{o,k}^{(s)}\}_{s=1}^{N}$ denote Gaussian measurements with $\hat{\mathbf{Z}}_{o,k}^{(s)} \sim \mathcal{N}(\mathbf{Z}_{o,k}^{(s)}+\mu^{(s)}, \Sigma^{(s)})$.
Then there exists a unique barycenter $\hat{\mathbf{Z}}_{o,k}^{(\text{WB})} \sim \mathcal{N}({\mathbf{Z}_{o,k}+\mu^{(\text{WB})}}, {\Sigma^{(\text{WB})}})$ with barycentric coordinates $\{\lambda_s\}_{s=1}^N$ \cite{mallasto2017}.
\end{lemma}
If the barycenter $\hat{\mathbf{Z}}_{o,k}^{(\text{WB})}$ is non-degenerate, then the mean $\mathbf{Z}_{o,k}+\mu^{(\mathrm{WB})}$ and covariance $\Sigma^{(\mathrm{WB})}$ satisfy

\begin{subequations}
\label{theorem_avg}
\begin{align}
{\mathbf{Z}_{o,k}+\mu^{(\text{WB})}} = \sum_{s=1}^N \lambda_s (\mathbf{Z}_{o,k}^{(s)}+\mu^{(s)}), \\
{\Sigma^{(\text{WB})}} = \sum_{s=1}^N \lambda_s \left({\Sigma^{(\text{WB})}} \Sigma^{(s)} {\Sigma^{(\text{WB})}} \right)^{1/2}.
\end{align}
\end{subequations}

Each weight of the sensors is defined as a positive scalar: 
$\lambda_1^{(\text{LiDAR})}$, $\lambda_2^{(\text{Camera})}$, 
$\lambda_3^{(\text{V2X})}$, with $\sum_{s=1}^{3} \lambda_s = 1$. 
Substituting (\ref{sensor_dis}) and the weights into (\ref{theorem_avg}) gives:

\vspace{-1.5em}

\begin{subequations}
\begin{align}
\mathbf{Z}_{o,k} + \mu^{(\text{WB})} 
&= \lambda_1^{(\text{LiDAR})} \left( \mathbf{Z}_{o,k}^{(\text{LiDAR})} + \mu^{(\text{LiDAR})} \right) \nonumber \\
&\quad + \lambda_2^{(\text{Camera})} \left( \mathbf{Z}_{o,k}^{(\text{Camera})} + \mu^{(\text{Camera})} \right) \nonumber \\
&\quad + \lambda_3^{(\text{V2X})} \left( \mathbf{Z}_{o,k}^{(\text{V2X})} + \mu^{(\text{V2X})} \right), \label{eq:weighted_mean} \\
\Sigma^{(\text{WB})} 
&= \lambda_1^{(\text{LiDAR})} 
\left( \Sigma^{(\text{WB})} \Sigma^{(\text{LiDAR})} \Sigma^{(\text{WB})} \right)^{1/2} \nonumber \\
&\quad + \lambda_2^{(\text{Camera})} 
\left( \Sigma^{(\text{WB})} \Sigma^{(\text{Camera})} \Sigma^{(\text{WB})} \right)^{1/2} \nonumber \\
&\quad + \lambda_3^{(\text{V2X})} 
\left( \Sigma^{(\text{WB})} \Sigma^{(\text{V2X})} \Sigma^{(\text{WB})} \right)^{1/2}. \label{eq:weighted_cov}
\end{align}
\end{subequations}
The obstacle position based on Wasserstein Barycenter from all sensors can be written as:
\begin{equation}
\label{WB_measurement}
\hat{\mathbf{Z}}_{o,k}^{(\text{WB})} \sim \mathcal{N}(\mathbf{Z}_{o,k} + \mu^{(\text{WB})}, \Sigma^{(\text{WB})}),
\end{equation}
where $\Sigma^{(\text{WB})}$ is the covariance of Wasserstein Barycenter.


\subsection{Wasserstein Barycenter-based Control Barrier Function}
In this subsection, we present the integration of the Wasserstein Barycenter into the construction of Control Barrier Functions (CBFs). The CBF at time step $k$ is defined as:
\begin{subequations}
\begin{align}
\label{cbf_original}
h({\mathbf{Z}}_{v,k}, {\mathbf{Z}}_{o,k}) &= \left\| 
{\mathbf{Z}}_{v,k}
- 
{\mathbf{Z}}_{o,k}
\right\|_2 - D, \\
D &= R_{v}+R_{o}+ d_{\text{s}},
\end{align}
\end{subequations}
where $R_{v}$ and $R_{o}$ denote the radius of the AV and the obstacle, and  $d_{\text{s}}$ is the safety distance between the AV and the obstacle.
(\ref{cbf_original}) considers the true positions of both the vehicle and the obstacle. However, in practical scenarios, only noisy state measurements and estimates are available. 
It should also be noted that the CBF should be formulated from two Gaussian measurements of the vehicle position and the Wasserstein Barycenter.
Hence, (\ref{cbf_original}) is reformulated using $I$ samples from the measurement distribution in (\ref{gps_measure}) and with $J$ samples from the Wasserstein Barycenter distribution in (\ref{WB_measurement}). Thus, the CBF formulation from sampling the two distributions is given by:
\begin{subequations}
\begin{gather}
h_{I \times J}(\hat{\mathbf{Z}}_{v,k}, \hat{\mathbf{Z}}_{o,k}^{\text{(WB)}}) = \left\| 
\hat{\mathbf{Z}}_{(v,k),i}
- 
\hat{\mathbf{Z}}_{(o,k),j}^{\text{(WB)}}
\right\|_2 - D, \label{cbf_measurement-a} \\
\quad i = 1, \ldots, I, \text{and} \quad j = 1, \ldots, J. \nonumber
\intertext{For notational simplicity, we denote the CBF in (\ref{cbf_measurement-a}) as $h(\cdot)$. The modified CBF optimization constraint based on sampling the Wasserstein Barycenter obstacle position and the vehicle position is given by}
 \text{CBC} \left(\hat{\mathbf{Z}}_{v,k}, \mathbf{u}, \hat{\mathbf{Z}}_{o,k}^{\text{(WB)}}\right) = L_f h(\cdot) 
  + L_g h(\cdot) \mathbf{u} 
  + \alpha\left( h(\cdot) \right) 
  \geq 0. \label{cbf_measurement-b}
\end{gather}
\end{subequations}
Note that the CBC in (\ref{cbf_measurement-b}) follows a distribution $\mathbb{P}$ and the next section discusses on how to compute a safe control action on this distribution.

\section{Risk-Aware Safety-Critical Control Formulation}
Due to the stochasticity in obstacle and vehicle position, the CBF in optimization formulation is considered a chance-constrained optimization problem that accommodates uncertainties. Thus, the chance-constraint is given by 
\begin{equation} \label{Chance-Constraint}
    \mathbb{P} \left( \text{CBC} \left(\hat{\mathbf{Z}}_{v,k}, \mathbf{u}, \hat{\mathbf{Z}}_{o,k}^{\text{(WB)}} \right) \geq 0 \right) \geq 1-\epsilon,
\end{equation}
where $\epsilon \in (0,1)$ is a user defined risk tolerance. For example, $\epsilon =0.05$ denotes $5\%$ risk tolerance. Note that this does not mean that the system is under 5\% risk, but implies that the $\epsilon$-percentile risk is optimized. For notional simplicity, from here on, the CBC in (\ref{cbf_measurement-b}) is denoted as $CBC(.)$. The constraint in (\ref{Chance-Constraint}) makes the optimization formulation non-convex and computationally intractable \cite{nemirovski2007convex}. 

However, \cite{rockafellar2000optimization} provides a convex CVaR approximation of chance constraints. We introduce the following risk-based Definition~\ref{VaR_definition}.

\begin{definition} \label{VaR_definition}
The Value at Risk (VaR) of the random variable $CBC$ under probability distribution $\mathbb{P}$ at risk level $\epsilon \in (0,1)$ is defined as
\begin{align*}
\text{VaR}_{\epsilon}^{\mathbb{P}}(CBC)
:=
\inf_{\gamma \in \mathbb{R}}
\left\{
\gamma \;|\;
\mathbb{P}(CBC \le \gamma) \ge \epsilon
\right\}.
\end{align*}
\end{definition}

The VaR corresponds to the inverse of the cumulative distribution function (CDF) at probability level $\epsilon$. 
Clearly, $\gamma \in \mathbb{R}$ represents a threshold value of the loss and does not provide much information about the distribution tail, which may still lead to intractable optimization. 
To better quantify tail risk while preserving convexity in optimization, the Conditional Value at Risk (CVaR) is adopted and defined in Definition~\ref{CVaR_definition}.
 
\begin{definition}\label{CVaR_definition}
The Conditional Value at Risk (CVaR) of the random variable $CBC$ under probability distribution $\mathbb{P}$ at risk level $\epsilon \in (0,1)$ is defined as
\begin{align*}
\text{CVaR}_{\epsilon}^{\mathbb{P}}\left( CBC(.) \right) \triangleq \\ \mathbb{E}_{\mathbb{P}} \left[ CBC(.) \;\middle| \right.
\left. CBC(.) \leq \text{VaR}_{\epsilon}^{\mathbb{P}}(CBC(.)) \right].
\end{align*}
\end{definition} 

The CVaR can be seen as the stochastic safety violation of the autonomous vehicle with VRUs. In the following part, it is shown that the CVaR can be written as a tractable convex function.
\begin{lemma} \cite{rockafellar2000optimization}\label{Convex-CVAR}
The approximated function, sampled from the distribution of CBC, given by
\begin{align}
F_{\epsilon}\big(\mathrm{CBC}(\cdot), \gamma\big)
= \gamma - \frac{1}{N \epsilon} \sum_{i=1}^{N} \big[ \gamma - \mathrm{CBC}(\cdot) \big]^+,
\end{align}
is convex.
\end{lemma}
\begin{pf}
The proof follows from \cite{rockafellar2000optimization}. Clearly,
\[
\frac{1}{N} \sum_{i=1}^{N} \big[\gamma - \mathrm{CBC}(\cdot)\big]^+
= \mathbb{E}_P\!\left[\,\gamma - \mathrm{CBC}(\cdot)\,\right]^+,
\]
and
\[
\big[ \gamma - \mathrm{CBC}(\cdot) \big]^+ = \max\{0,\, \gamma - \mathrm{CBC}(\cdot)\}.
\]
Since $\max\{0,\,x\}$ is convex and $\gamma$ is constant, $\max\{0,\, \gamma-\mathrm{CBC}(\cdot)\}$ is convex whenever $\mathrm{CBC}(\cdot)$ is convex. Expectations preserve convexity, hence \eqref{Convex-CVAR} is convex.
\end{pf}
\begin{theorem} \label{cvar_cbc_thm}
The CVaR of the CBC is determined by
\begin{equation} \label{CVAR-CBC}
    \min_{\gamma \in \mathbb{R}} F_{\epsilon}\big( CBC(.), \gamma \big) .
\end{equation}
\end{theorem}
\begin{pf}
    The proof is inspired from \cite{rockafellar2000optimization}. From Definition~\ref{VaR_definition} and Lemma~\ref{Convex-CVAR}, it is evident that the values of $\gamma$ that give the minimum of $F_{\epsilon}\big(CBC(.), \gamma \big) $, is $\text{VaR}_{\epsilon}^{\mathbb{P}}(CBC(.))$. Thus, 
\begin{align*}
\min_{\gamma \in \mathbb{R}} &\; F_{\epsilon}\big( CBC(.), \gamma \big)= F_{\epsilon}\big( CBC(.), \text{VaR}_{\epsilon}^{\mathbb{P}}(CBC(.)) \big) \\
= &\; \text{VaR}_{\epsilon}^{\mathbb{P}}(CBC(.)) + \\
&\frac{1}{N \epsilon} \sum_{i=1}^{N}  \big[ CBC(.) - \text{VaR}_{\epsilon}^{\mathbb{P}}(CBC(.)) \big]^+.
\end{align*}
\end{pf}
From the definition of CVaR, this can be approximated as
\begin{align*}
    \approx \text{VaR}_{\epsilon}^{\mathbb{P}}(CBC(.)) + \text{CVaR}_{\epsilon}^{\mathbb{P}}(CBC(.)) \\
    - \text{VaR}_{\epsilon}^{\mathbb{P}}(CBC(.)) = \text{CVaR}_{\epsilon}^{\mathbb{P}}(CBC(.)).
\end{align*}
\begin{corollary} \label{CVaR_collary}
Minimizing the safety violations is equivalent to minimizing $F_{\epsilon}\big( CBC(.), \gamma \big) $ over all $(\mathbf{u},\gamma) \in \mathcal{U} \times \mathbb{R}$
\begin{align} \label{CVAR-CBC-min}
    \min_{\mathbf{u} \in \mathcal{U}} \text{CVaR}_{\epsilon}^{\mathbb{P}}\left(CBC(.)\right) = \min_{(\mathbf{u},\gamma) \in \mathcal{U}\times \mathbb{R}} F_{\epsilon}\big( CBC(.), \gamma \big).
\end{align}
\begin{itemize}
    \item  $F_{\epsilon}\big( CBC(.), \gamma \big)$ can be minimized jointly over the variables $(\mathbf{u}, \gamma)$ through convex optimization 
    \item The pair $(\mathbf{u}^*, \gamma^*)$ achieves joint minimum iff $\mathbf{u^*}$ minimizes CVaR and $\gamma^*$ is the VaR associated with $\mathbf{u^*}$.
\end{itemize}
\end{corollary}
Corollary~\ref{CVaR_collary} intuitively shows that it is much simpler to work on the convex formulation (\ref{Convex-CVAR}) with respect to $(\mathbf{u}, \gamma)$ than the direct optimization of VaR, which could be intractable. 
The definitions used for VaR, CVaR, and the above shown proofs differ from the literature as here we directly identify the value at risk based on the $\epsilon$-percentile of the distribution. This implies that the following constraints are equivalent.
\[
    \mathbb{P}\{CBC(.)\leq \gamma\} \geq \epsilon \leftrightarrow \mathbb{P}\{CBC(.)\geq \gamma\} \geq 1-\epsilon \]\[
     \leftrightarrow \text{VaR}_{\epsilon}^{\mathbb{P}}(CBC(.))\leq \gamma.
\]
In short, we identify the tail of the distribution that contributes to the worst-case risk and formulate the optimization problem to minimize it.

\subsection{WB-CVaR-CBF Formulation}
This subsection shows the optimization problem (OP) formulation: Wasserstein Barycenter - Conditional Value at Risk - Control Barrier Function Optimization Problem (WB-CVaR-CBF) and discusses the feasibility of the optimization problem. From the QP formulation (\ref{cbf_qp}), we have
\begin{equation} \label{WB-CVaR-CBF-OP}
\begin{aligned}
    \mathbf{u}_{\text{safe}} =\; & \underset{\mathbf{u} \in \mathcal{U},\, \gamma \in \mathbb{R}}{\text{min}} \; \frac{1}{2} \left\| \mathbf{u} - \mathbf{u}_{\text{nom}} \right\|^2, \\
     \text{s.t.} \; \gamma - & \frac{1}{N\epsilon} \sum_{i=1}^{N} \left[ \gamma - \text{CBC} \left(\hat{\mathbf{Z}}_{v,k}, \mathbf{u}, \hat{\mathbf{Z}}_{o,k}^{\text{(WB)}} \right) \right]^+ \geq 0.
\end{aligned}
\end{equation}
\begin{flushright}
    (WB-CVaR-CBF)
\end{flushright}
In the remainder of this section, we discuss the specific assumptions under which this OP formulation works. 

\begin{assumption} \label{assumption}
The camera, LiDAR, and V2X identify and track the same object. This assumption is quite valid from the perspective of the controller, that the sensor fusion, sensing, and tracking algorithms provide this information \cite{huang2023v2x}. 
\end{assumption}
The feasibility of our WB-CVaR-CBF formulation is implied through Assumption~\ref{assumption}, Lemma~\ref{wb_unique}, Lemma~\ref{Convex-CVAR}, and Theorem~\ref{cvar_cbc_thm}. Lemma~\ref{wb_unique} shows that a unique barycenter always exists. So, with Assumption~\ref{assumption}, the control barrier constraint (\ref{cbf_measurement-b}) always exists. The optimization problem in (\ref{WB-CVaR-CBF-OP}) is feasible through convex optimization, as shown in Corollary~\ref{CVaR_collary}. So, the solution is feasible as long as $CBC(.)$ is convex. 

The optimization problem can become infeasible when the AV is very close to the obstacle. This can be the case for late detections, suddenly appearing obstacles, and AVs or obstacles at high speeds, which are rare and extreme scenarios and are not considered in this work. 

\section{Experimental and Numerical Example: Autonomous Vehicle Control} \label{Numerical_Example}
\subsection{Experimental Setup}

\begin{figure}[h]
      \centering
      \includegraphics[scale=0.4]{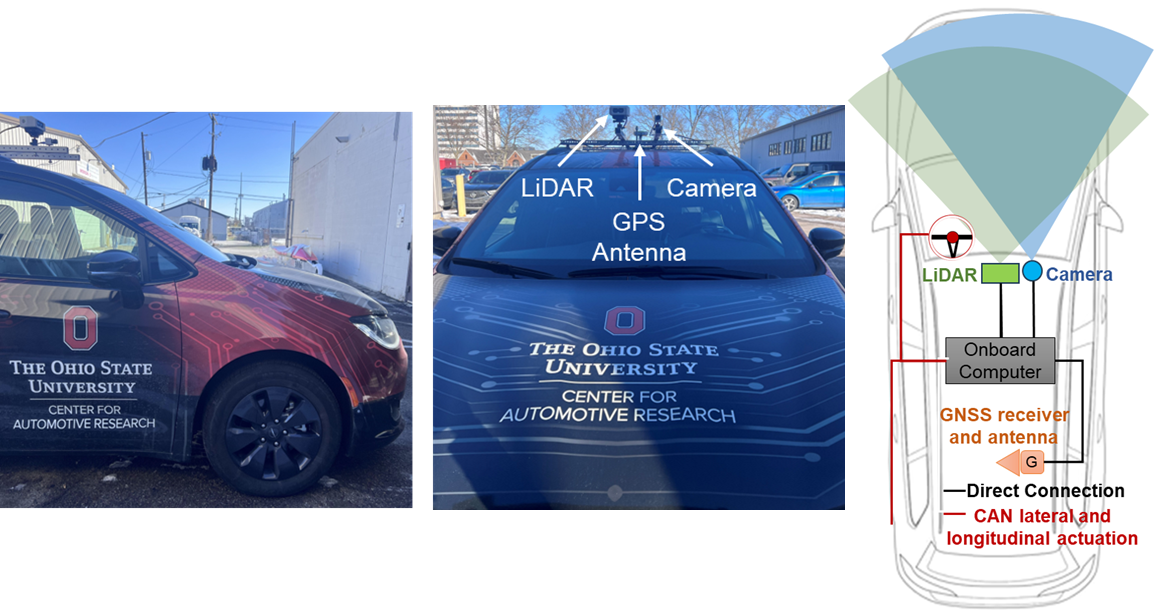}
      \caption{ The experimental vehicle and the sensors mounted position.}
      \label{experiment_vehicle}
\end{figure}

The proposed framework is assessed using both experimental trials and numerical simulations. Experimentally, we demonstrate on an AV that the controller can be deployed in real time and successfully avoid moving obstacles. The test vehicle is shown in Fig. \ref{experiment_vehicle}, whose setup follows the protocol in \cite{Renganathan2025Experimental}. To evaluate robustness under varying noise levels, we performed Monte Carlo simulations with 100 runs per scenario. In both the experimental and simulation settings, the AV and vulnerable road users (VRUs) follow predefined trajectories that intersect such that the VRU enters the AV’s path within a 10 m radius. The AV is expected to detect the VRU and execute a safe collision-avoidance maneuver.
\subsubsection{Scenarios and Parameters}
Evaluations primarily focus on the controller's performance under varying levels of uncertainty. Specifically, we validate the controller in the following three scenarios:
\begin{itemize}
    \item Scenario 1: The AV receives low-uncertainty GPS data, and all three sensors operate under low-noise conditions.  
    \item Scenario 2: The AV receives high-uncertainty GPS data, while the remaining sensors still operate with low noise.
    \item Scenario 3: Both GPS and V2X measurements are affected by high levels of uncertainty and noise.
\end{itemize}

The optimization problems are solved using IPOPT solver in CasADi framework under the same set of parameters as shown in Table \ref{tab:parameters} to ensure result comparability. 
\begin{table}[h!] 
    \centering
    \caption{Parameters}
    \label{tab:parameters}
    \begin{tabular}{|c|c|} 
        \hline
        \textbf{Vehicle parameter and Constraints} & \textbf{Value} \\ \hline 
        Safety Distance & $3 m$\\ \hline
        Vehicle Radius & $1.8 m$\\ \hline
        obstacle (VRU) Radius & $1 m$\\ \hline
        obstacle (VRU) Speed & $4.5 m/s$\\ \hline
        $\lambda_{\text{LiDAR}}, \lambda_{\text{Camera}}, \lambda_{\text{V2X}}$ & $0.4, 0.4, 0.2$\\ \hline
    \end{tabular}
\end{table}
\vspace{-1.5em}

\subsubsection{Comparative Analysis}
For comparison, all trajectory-following controllers use MPC as the nominal controller and differ only in the safety filter:
\begin{itemize}
\item \textbf{CBF:} The CBF constraint in \eqref{cbf_qp} is enforced along the entire trajectory using the mean of the three sensor measurements as the obstacle estimate:
\begin{equation}
h(\hat{\mathbf{Z}}_{v,k}, \hat{\mathbf{Z}}_{o,k}^{(\text{mean})}) =
\|\hat{\mathbf{Z}}_{v,k} - \hat{\mathbf{Z}}_{o,k}^{(\text{mean})}\|_2 - D .
\end{equation}

\item \textbf{WB-CBF:} The same CBF constraint is used, but the obstacle position is estimated using the WB of the sensor measurements:
\begin{equation}
h(\hat{\mathbf{Z}}_{v,k}, \hat{\mathbf{Z}}_{o,k}^{(\text{WB})}) =
\|\hat{\mathbf{Z}}_{v,k} - \hat{\mathbf{Z}}_{o,k}^{(\text{WB})}\|_2 - D .
\end{equation}
\end{itemize}

\subsubsection{Metrics}
To evaluate performance over all $N_t$ test cases, we compute the following metrics. For all metrics except the success rate, averages are taken only over the successful runs, $N_s$.

\begin{itemize}
    \item \textbf{Success Rate (SR):} A run is deemed successful if the minimum vehicle--obstacle separation along the trajectory is always greater than the required clearance $d_{\min}=2.8\,\mathrm{m}$ (vehicle radius $1.8\,\mathrm{m}$ + obstacle radius $1.0\,\mathrm{m}$). The success rate is $\mathrm{SR}=N_s/N_t$.

    \item \textbf{Minimum Distance to obstacle (MDP):} For each successful run, let $d_{m,i}$ denote the minimum distance to the nearest obstacle during avoidance. We report the mean over successful runs,
    \[
        \mathrm{MDP}=\frac{1}{N_s}\sum_{i=1}^{N_s} d_{m,i}.
    \]
    An ideal clearance of at least $5.8\,\mathrm{m}$ (vehicle radius $1.8\,\mathrm{m}$ + obstacle radius $1.0\,\mathrm{m}$ + $3.0\,\mathrm{m}$ safety buffer) satisfies the desired safety margin. Values below $5.8\,\mathrm{m}$ indicate behavior that prioritizes reference-path tracking over conservative separation.
\end{itemize}

\subsection{Validation Results}
\begin{figure}[h]
      \centering
      \includegraphics[scale=0.37]{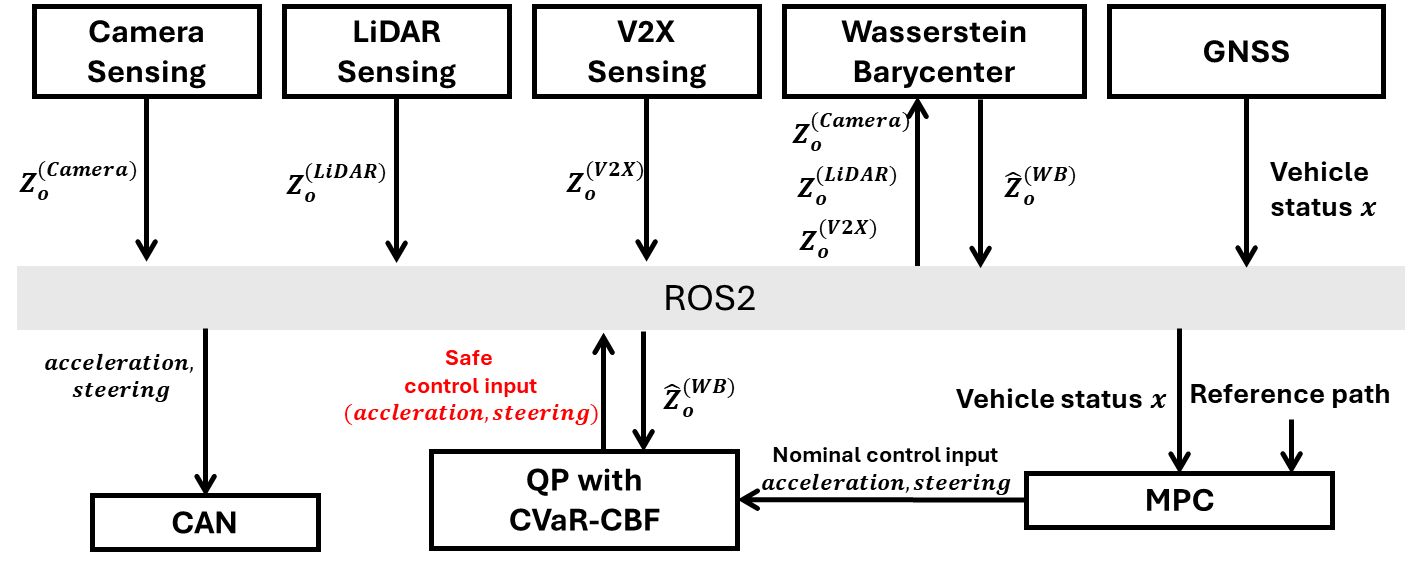}
      \caption{On-vehicle software architecture and data flow.}
      \label{experiment_vehicle_framework}
\end{figure}

The controller stack is implemented in Python and deployed as a set of ROS~2 nodes (Fig.~\ref{experiment_vehicle_framework}). Sensing nodes (Camera, LiDAR, V2X, and GNSS) publish raw and preprocessed measurements on ROS~2 topics. A fusion node computes the Wasserstein barycenter of these sensor distributions to obtain a unified estimate of obstacle states, which is then re-published for downstream modules.
The nominal trajectory-following command $\mathbf{u}_{\text{nom}}$ is generated by an MPC node. A safety-filter node subscribes to the fused obstacle state, and at each control cycle solves a QP with CVaR–CBF constraints to minimally modify $\mathbf{u}_{\text{nom}}$, producing the safe command $\mathbf{u}_{\text{safe}}=\{\text{acceleration},\text{steering}\}$. The resulting control input is passed to a CAN bus node and transmitted to the vehicle actuators via the CAN bus. All inter-module communication is handled through ROS~2 topics and services.
\begin{figure}[h]
      \centering
      \includegraphics[scale=0.8]{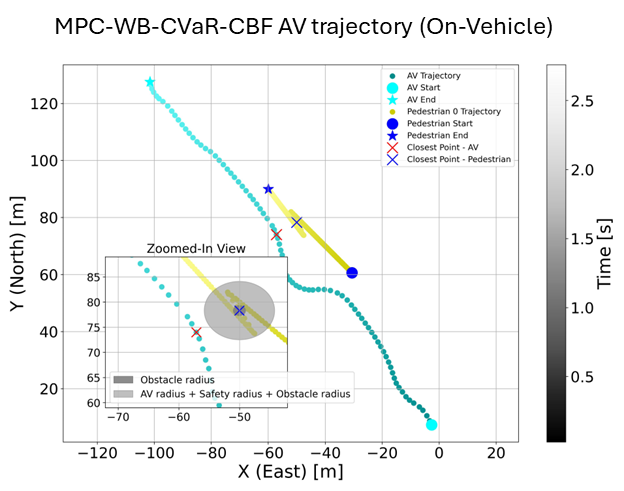}
      \caption{On-vehicle trajectory with MPC-WB-CVaR-CBF under scenario 2.}
      \label{experiment_vehicle_trajectory}
\end{figure}

\begin{figure*}
      \centering
      \includegraphics[scale=0.55]{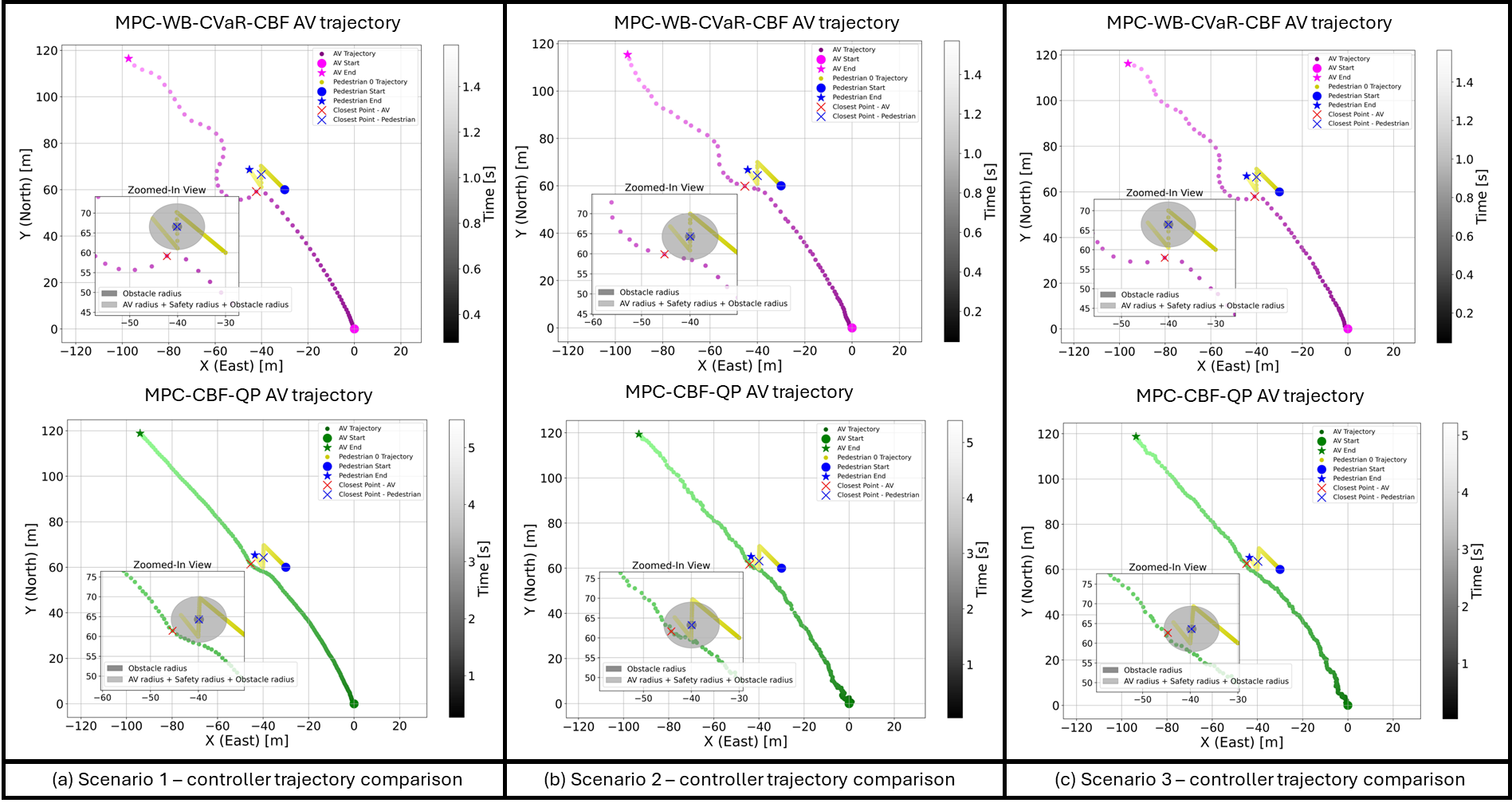}
    \caption{Trajectories under MPC–WB–CVaR–CBF (proposed) and MPC–CBF–QP (baseline). (a) Scenario 1: both safe; (b) Scenario 2: proposed safe, baseline unsafe; (c) Scenario 3: proposed safe, baseline unsafe.}
      \label{trajectory_figs}
\end{figure*}

The trajectory under Scenario~2 using the proposed controller is shown in Fig.~\ref{experiment_vehicle_trajectory}. A numerical simulation with the same sensor noise setting is provided in Fig.~\ref{trajectory_figs} (b) scenario 2. These results demonstrate that the proposed control framework is deployable on a full-scale autonomous vehicle and can successfully avoid a moving obstacle. The remaining discrepancy in tracking performance between experiment and simulation primarily stems from the bicycle model vehicle dynamics used in the simulator.


\subsection{Numerical Testing Results}
Having validated on-vehicle implementation, we assess robustness to sensor noise through a Monte Carlo study. The next section details 100 runs numerical experiments for each scenario.

\subsubsection{Scenario 1:}
This scenario has low GPS  uncertainty and low obstacle detection noise, and this is close to an ideal scenario. The GPS operates with $\sigma^{(\text{GPS})} = 0.1$, LiDAR with $\sigma^{(\text{LiDAR})} = 0.1$, camera with $\sigma^{(\text{Camera})} = 0.2$, and $\sigma^{(\text{V2X})} = 1.0$. The V2X is perceived to have higher noise due to transmission delays. The results for this case are summarized in Table-\ref{tab:s1_results}. In this setting, both MPC-CBF and MPC-WB-CVaR-CBF reach 100\% safety in all trials. The WB-CBF achieves 90\% success rate, suggesting that additional fusion provides limited benefit when the sensing information is reliable. The trajectory comparison between MPC-CBF and MPC-WB-CVaR-CBF is illustrated in Fig.~\ref{trajectory_figs}(a). The WB-CBF trajectories are omitted for brevity since their behavior is similar to the MPC-CBF controller in this scenario.
\begin{table}[h]
\centering
\caption{obstacle interaction results over 100 runs in Scenario 1; $\mu^{(\text{GPS})}=0,\ \sigma^{(\text{GPS})}=0.1$.}
\label{tab:s1_results}
\small
\begin{tabular}{lcc}
\toprule
\multirow{2}{*}{Method} &
\multicolumn{2}{l}{\parbox{0.6\linewidth}{
$ \mu^{(\mathrm{LiDAR})}=0,\ \sigma^{(\mathrm{LiDAR})}=0.1;$\\
$\mu^{(\mathrm{Camera})}=0,\ \sigma^{(\mathrm{Camera})}=0.2;\\ 
\mu^{(\mathrm{V2X})}=0,\ \sigma^{(\mathrm{V2X})}=1.0$.}} \\
\cmidrule(l){2-3}
& SR (\%) & MDP (m) \\
\midrule
CBF            & 100\%          &  6.15 \\
WB-CBF            & 90\%          &  5.64\\
\rowcolor{gray!15}
WB-CVaR-CBF    & \textbf{100\%}    & 7.03  \\
\bottomrule
\end{tabular}
\end{table}
\vspace{-1.5em}

\subsubsection{Scenario 2:}
In this scenario, GPS uncertainty is attributed to a highly disturbed environment, such as GPS attenuation. The GPS noise level is assumed to be $\sigma^{(\text{GPS})} = 0.5$, corresponding to a 0.5-meter error. Meanwhile, the obstacle positions detected by the sensors remain at a low noise level, as discussed in the previous scenario. The results are presented in Table~\ref{tab:s2_results}. 

Under this condition, the MPC-CBF controller encounters a noticeable performance degradation, with only 67\% SR. Incorporating WB fusion improves the SR to 75\%, indicating that multi-modal fusion helps mitigate the impact of increased localization uncertainty. By further introducing the CVaR-based risk-aware constraint, the proposed MPC-WB-CVaR-CBF achieves significantly better performance with 97\% SR. This result highlights the benefit of combining distributional fusion with risk-aware control under elevated sensing uncertainty. Fig.~\ref{trajectory_figs}(b) illustrates the resulting trajectories, where the proposed controller maintains larger safety margins relative to the baseline.
\begin{table}[h]
\centering
\caption{obstacle interaction results over 100 runs in Scenario 2; $\mu^{(\text{GPS})}=0,\ \sigma^{(\text{GPS})}=0.5$.}
\label{tab:s2_results}
\small
\begin{tabular}{lcc}
\toprule
\multirow{2}{*}{Method} &
\multicolumn{2}{l}{\parbox{0.6\linewidth}{
$ \mu^{(\mathrm{LiDAR})}=0,\ \sigma^{(\mathrm{LiDAR})}=0.1;$\\
$\mu^{(\mathrm{Camera})}=0,\ \sigma^{(\mathrm{Camera})}=0.2;\\ 
\mu^{(\mathrm{V2X})}=0,\ \sigma^{(\mathrm{V2X})}=1.0$.}} \\
\cmidrule(l){2-3}
& SR (\%) & MDP (m) \\
\midrule
CBF            & 67\%          &  6.43 \\
WB-CBF            & 75\%          &  5.07 \\
\rowcolor{gray!15}
WB-CVaR-CBF    & \textbf{97\%}         & 6.93 \\
\bottomrule
\end{tabular}
\end{table}
\vspace{-1.5em}

\subsubsection{Scenario 3:}
This scenario considers extreme V2X conditions where signals are affected by transmission delays or disturbances. The GPS uncertainty remains the same as in Scenario~2, introducing input disturbance, while V2X measurements contain bias and noise with $\mu^{(\text{V2X})}=-1$ and $\sigma^{(\text{V2X})}=1.0$. LiDAR and camera detections maintain low noise as in previous cases.

The results are shown in Table~\ref{tab:s3_results} and Fig.~\ref{trajectory_figs}(c). Under these conditions, MPC-CBF results in an 8\% unsafe rate due to biased V2X measurements. WB-CBF shows comparable performance (90\% SR), indicating that WB fusion alone does not fully mitigate biased sensing. In contrast, the proposed MPC-WB-CVaR-CBF achieves 100\% safety by explicitly accounting for sensing uncertainty and tail risk.
\begin{table}[h]
\centering
\caption{obstacle interaction results over 100 runs in Scenario 3; $\mu^{(\text{GPS})}=0,\ \sigma^{(\text{GPS})}=0.5$.}
\label{tab:s3_results}
\small
\begin{tabular}{lcc}
\toprule
\multirow{2}{*}{Method} &
\multicolumn{2}{l}{\parbox{0.6\linewidth}{
$ \mu^{(\mathrm{LiDAR})}=0,\ \sigma^{(\mathrm{LiDAR})}=0.1;$\\
$\mu^{(\mathrm{Camera})}=0,\ \sigma^{(\mathrm{Camera})}=0.2;\\ 
\mu^{(\mathrm{V2X})}=-1,\ \sigma^{(\mathrm{V2X})}=1.0$.}} \\
\cmidrule(l){2-3}
& SR (\%) & MDP (m) \\
\midrule
CBF            & 92\%          & 6.535  \\
WB-CBF            & 90\%          & 5.56  \\
\rowcolor{gray!15}
WB-CVaR-CBF    & \textbf{100\%}         &  7.168 \\
\bottomrule
\end{tabular}
\end{table}

\section{Conclusion}

This paper presented a risk-aware safe control framework for autonomous driving under localization and multi-modal sensing uncertainty. 
By integrating WB fusion with a CVaR-CBF safety filter, the proposed architecture systematically incorporates noisy obstacle estimates into safety-critical control. 
The on-vehicle obstacle-crossing experiment validated WB-CVaR-CBF with MPC on a full-scale AV. It supports that combining multi-modal sensor fusion with risk-aware safety filtering provides a practical and effective way to improve safety under uncertain sensing conditions.
The proposed method was also evaluated in three uncertainty scenarios, including low-noise sensing, elevated GPS uncertainty, and biased V2X measurements.
Across all scenarios, it consistently outperformed the baseline methods.
The benefit of the proposed framework becomes more robust as uncertainty and bias increase. 

For future work, signal loss can be considered, such as packet loss in V2X-assisted sensing. Although this issue is not modeled in the present paper, the proposed WB fusion could be extended by adaptively adjusting the barycentric weights according to sensor availability. Investigating the extension would further improve robustness in multi-modal sensing environments.

\bibliography{ifacconf_v2}                 

\end{document}